\documentclass{aa}  

\usepackage{txfonts}

\usepackage{graphicx}
\usepackage{amssymb}
\usepackage{epstopdf}
\usepackage{lscape}
\usepackage{rotating}
\usepackage{placeins}

\usepackage{natbib,twoopt}
\usepackage[breaklinks=true]{hyperref} 
\bibpunct{(}{)}{;}{a}{}{,} 
\makeatletter
\newcommandtwoopt{\citeads}[3][][]{\href{http://adsabs.harvard.edu/abs/#3}%
{\def\hyper@linkstart##1##2{}%
\let\hyper@linkend\@empty\citealp[#1][#2]{#3}}}
\newcommandtwoopt{\citepads}[3][][]{\href{http://adsabs.harvard.edu/abs/#3}%
{\def\hyper@linkstart##1##2{}%
\let\hyper@linkend\@empty\citep[#1][#2]{#3}}}
\newcommandtwoopt{\citetads}[3][][]{\href{http://adsabs.harvard.edu/abs/#3}%
{\def\hyper@linkstart##1##2{}%
\let\hyper@linkend\@empty\citet[#1][#2]{#3}}}
\newcommandtwoopt{\citeyearads}[3][][]%
{\href{http://adsabs.harvard.edu/abs/#3}
{\def\hyper@linkstart##1##2{}%
\let\hyper@linkend\@empty\citeyear[#1][#2]{#3}}}
\makeatother

\def\ERFI{\,{\rm erfi}}
\def\KPC{{\rm kpc}}
\def\PC{{\rm pc}}
\def\KPS{{\rm km}\,{\rm s}^{-1}}
\def\KMS2{{\rm km}^2\,{\rm s}^{-2}}
\def\MSUN{{\rm M}_\odot}
\def\MSUNSPC{\MSUN\PC^{-2}}
\def\MSUNCPC{\MSUN\PC^{-3}}
\def\mMSUNCPC{{\rm m}\MSUN\PC^{-3}}

\def\MU{\mu{\rm m}}

\def\PM#1#2{^{+#1}_{-#2}}
\def\V{{\rm v}}
\def\CITE#1{\citeyearads{#1}}

\def\DETAILS#1{} 
\usepackage{color}

\begin{document}

\title{The difficulty of measuring the local dark matter density}
\author{Frederic V. Hessman} 
\institute{Institut f\"ur Astrophysik, Georg-August-Universit\"at, Friedrich-Hund-Platz 1, 37077 G\"ottingen, Germany}
\date{Received March 5, 2015; accepted May 18, 2015}

\abstract{
The analysis of the vertical velocity dispersion of disc stars in the local Milky Way is the most direct astronomical means of estimating the local dark matter density, $\rho_{DM}$.
Current estimates for $\rho_{DM}$ based on the mid-plane dynamic density use a local baryonic correction that ignores the non-local effects of spiral structure and significantly underestimates the amount of dynamically relevant gas now known to be present in the ISM;
the additional gas plus the remaining uncertainties make it practically impossible to measure $\rho_{DM}$ from mid-plane kinematics alone.
The sampling of inhomogeneous tracer populations with different scale-heights and scale-lengths results in a systematic increase in the observed dispersion gradients and changes in the nominal density distributions that, if not properly considered, can be misinterpreted as a sign of more dark matter.
If the disc gravity is modelled locally using an infinite disc, the local variation in the vertical gravity due to the globally exponential disc components results in an underestimation of the baryonic contribution by as much as $\sim$40\%.
Given only the assumptions of stationarity, an axially and vertically symmetric disc, doubly exponential tracer and mass-component density profiles, a phenomenologically justified model for the cross-dispersion component $\sigma_{Rz}$, and a realistic model for $g_z(z)$, it is possible to solve the full vertical Jeans equation analytically for the vertical dispersion $\sigma_{z}(z)$ and hence test the robustness of previous attempts at measuring $\rho_{DM}$.
When the model parameters for $\sigma_{Rz}$ are estimated from SEGUE G dwarf star data, it is still not possible to explain the difference in behaviour seen in the simple thick- and thin-disc datasets reported by Buedenbender et al. (2014).
Rather than being a fundamental problem with the kinematical model, this effect appears to be a further  sign of the difficulty of defining and handling kinematically homogeneous tracer populations.
}

\keywords{Galaxy: disc, Galaxy: kinematics and dynamics, Galaxy: solar neighbourhood, Cosmology: dark matter}

\maketitle


\section{Introduction}

Although there are many indirect lines of evidence for the presence of non-baryonic matter in the Universe, nowhere is a concrete estimate for the density of such dark matter (DM) more important than in the Solar neighbourhood: the local density determines how easily such matter might be detected in laboratory experiments.
Starting with the classic analyses of Kapteyn (\CITE{1922ApJ....55..302K}) and Oort (\CITE{1932BAN.....6..249O}, \CITE{1960BAN....15...45O}),
there have been many attempts to use the kinematics of stars in the local Milky Way (MW) to estimate the local density and distribution of mass.
Recently, the data required for such an analysis have become much better thanks to massive kinematic surveys like 
SEGUE (Yanny et al. \CITE{2009AJ....137.4377Y}),
RAVE (Steinmetz et al. \CITE{2006AJ....132.1645S}), and
APOGEE (Allende Prieto et al. \CITE{2008AN....329.1018A}).
Read (\CITE{2014JPhG...41f3101R}) has recently given a detailed review of the theoretical expectations from a cosmological/extragalactic perspective and the history of the astronomical attempts to measure the local DM mass density; he shows that recent estimates have yielded values in the range $\rho_{DM}\!=\!5-15\,\mMSUNCPC$, but that systematic effects are still the dominant source of error.\footnote{I use the astronomical unit $1\,\mMSUNCPC \equiv 0.001\,\MSUNCPC$ = 0.038\,GeV\,cm$^{-3}$ to avoid having to count the many zeros present when using $\MSUNCPC$.}
The measurement can be made using one of two different basic approaches, each with its own strengths and weaknesses.

One can attempt to fit the global kinematics of the stars and gas in the MW in detail and, by generating a global mass model, also estimate the local DM density (e.g.
Bienaym\'e, Robin \& Cr\'ez\'e \CITE{1987A&A...180...94B},
McMillan \CITE{2011MNRAS.414.2446M}).
The problem with this approach is that basic properties like the distances to kinematic structures (e.g. Binney \& Merrifield \CITE{1998gaas.book.....B}, Section 9.1.1; Reid et al. \CITE{2014ApJ...783..130R})
or even the distance to the center of the MW (e.g. Gillessen et al. \CITE{2009ApJ...692.1075G}; Sofue et al. \CITE{2011PASJ...63..867S}; Branham \CITE{2014Ap&SS.353..179B}) are difficult to measure; the area within which a reasonably robust analysis can be made is only between the end of the central bar and the Sun's Galactocentric radius, $R_0$ (Dehnen \& Binney \CITE{1998MNRAS.294..429D});
and the parameters of arbitrary models for the DM geometry must be fit, e.g. using a Navarrow-Frenck-White (\CITE{1997ApJ...490..493N})
or Einasto (\CITE{1965TrAlm...5...87E}) profile plus assumptions about the halo's oblateness.
A recent example of this approach is that of Piffl et al. (\CITE{2014MNRAS.445.3133P}): using $\sim\!200,000$ stars from the RAVE survey with Galactic heights $|z|\!<\!1.5\,\KPC$ and a 33-parameter model 
for the global mass-distribution and its effects on the stellar kinematics via assumed stellar distribution functions (DF), they derived a local DM density of $15\pm 2\,\mMSUNCPC$.
When kinematic data from the GAIA satellite become available, this method will be the only way of dealing with the complexity of a global data sample from a realistic MW (Binney \CITE{1998gaas.book.....B}) and the sheer enormity of the dataset will constrain the otherwise arbitrary model parameters to conform to reality.

An alternate approach better suited to the limits of current data and our knowledge of the structure of the MW (Carraro \CITE{2015arXiv150203151C}) is to try to measure the DM density locally using local stellar kinematics (e.g. 
Bahcall \CITE{1984ApJ...287..926B}, Kuijken \& Gilmore \CITE{1989MNRAS.239..571K}, Holmberg \& Flynn \CITE{2004MNRAS.352..440H}, Bovy \& Rix \CITE{2013ApJ...779..115B}). 
The advantage of this approach is that the proper motions and distances to all stars and hence the full three-dimensional densities, velocities, and dispersions can principally be measured, at least at large distances from the plane of the MW where the effects of DM should show up most directly, and the large-scale distribution of DM presumedly plays a minor role (e.g. Siebert et al. \CITE{2008MNRAS.391..793S}).
Here too the complex potential and DF models can be used, but they may not be able to ensure that the result is accurate (versus merely precise).
For example, the simple two-component St\"ackel potentials used by Bienyam\'e et al. (\CITE{2014A&A...571A..92B}) to model the disc and DM halo give the local disc a power-law rather than an exponential radial profile; any inaccuracies in the resulting model dispersion will be compensated by modifying the ill-constrained properties of the DM halo.

It is important to try to estimate the gravitational effects attributed to $\rho_{DM}$ locally on length-scales smaller than $R_0$, since there are competing theories like MOND (e.g. Milgrom \citeyearads{2014MNRAS.437.2531M}) that invoke changes in the behaviour of gravity on scales of whole galaxies -- the same scales relevant for studies of the MW's rotation curve.
If the local effects of DM in the MW can only be determined using the MW's rotation curve, we will have lost the information at one scale-length and have to deal with much more complexity and intrinsic uncertainties.
The purpose of this paper is then to probe how robust truly local estimates for $\rho_{DM}$ can be.
I first show that the extent to which the derived values of $\rho_{DM}$ are compromised by the many uncertainties and pitfalls in the methods and data used is generally not sufficently recognized.
Then, I describe a relatively simple method with which one can principally make more robust estimates for the local DM density using the full vertical Jeans equation; as few assumptions as possible are made about the large-scale stellar kinematics, MW properties, and unknown DM model parameters, and the effects of radial coupling and a realistic rather than an idealized toy model is used for the local gravity.
Finally, I apply this method to a particular dataset in order to see whether the challenges fundamentally lie in our ability to model the physics or to have adequate data.
As we shall see, the problems with many previous $\rho_{DM}$ estimates are serious and one must more carefully consider the inadequacies of models and the difficulty of defining and handling appropriate tracer datasets.


\section{Near-midplane estimates for the DM density}

\label{sec:midplane}

If the stellar motions and the local potential $\Phi(z)$ near the MW mid-plane are simple enough, the Jeans theorem says that $\V_z^2/2+\Phi$ is an approximate isolating integral and the dynamic local density $\rho_{dyn}(0)$ can be calculated from the distribution of mid-plane vertical velocities $\V_z(0)$ and the observed density distribution $\nu(z)$ of a tracer population (Kuijken \& Gilmore \CITE{1989MNRAS.239..571K}; see Sect.\,4.9.3 in Binney \& Tremaine \CITE{2008gady.book.....B}, hereafter BT08).  
Given enough good data, this method should yield a very robust estimate for $\rho_{dyn}(0)$ and -- given a sufficiently accurate model for the baryonic contributions -- a similarly robust estimate for the remaining $\rho_{DM}$.
This analysis has been made using many datasets and models (e.g. Kuijken \& Gilmore \CITE{1989MNRAS.239..651K}; Holmberg \& Flynn \CITE{2000MNRAS.313..209H}; Creze et al.\,\CITE{1998A&A...329..920C}; Garbari et al. \CITE{2012MNRAS.425.1445G}).

The problem with the conversion of the dynamic density, once measured, to the local DM density is the difficulty of estimating the baryonic contribution $\rho_{dyn,b}(0)$.
The observed mid-plane density of stars is fairly simple to estimate from local surveys: Read's (\CITE{2014JPhG...41f3101R}) re-tabulation of the Flynn et al.\,(\CITE{2006MNRAS.372.1149F}) model has $37.9$ and $3.5\,\mMSUNCPC$ from the thin and thick discs, respectively (if one assigns the local giants, white dwarfs, and brown dwarfs to the thin disc for this purpose).
If one assumes that the density-ratio of the thick disc to thin disc is the value measured photometrically 
by Juri\'c et al.\,(\CITE{2008ApJ...673..864J}), then the contributions have to be redistributed as $37.0$ and $4.4\,\mMSUNCPC$.
The local surface density of molecular, neutral, and warm gas has been classically estimated as $13\,\MSUNSPC$ by Holmberg \& Flynn (\CITE{2000MNRAS.313..209H}), from which one can estimate the local densities using typical observed vertical scale-heights and/or linewidths.
The stellar densities are generally thought to be good to about 10\%, whereas the gas densities are difficult to estimate; the latter are given errors of $\sim$50\% by Homberg \& Flynn (\CITE{2000MNRAS.313..209H}), suggesting a total error of about $25\,\mMSUNCPC$.
It should be noted that the local baryonic mass-density within this traditional model is dominated by the gas -- i.e. by the least well-determined component -- and the uncertainty is already much larger than the remaining $5\!-\!15\,\mMSUNCPC$ thought to be due to DM.

The total estimated baryonic density $\rho_b(0) \approx 91\,\mMSUNCPC$ can be converted to total surface density by adding up the contributions of individual kinematically isothermal components and hence depends upon the assumed or fitted potential.
Using the potential from Holmberg \& Flynn (\CITE{2004MNRAS.352..440H}), 
one obtains the values shown in Table\,1.
Like Read, I have used updated HI dispersions (Kalberla \& Dedes \CITE{2008A&A...487..951K}).
These surface density estimates are not internally-consistent -- one would have to repeat the fit of the K giant stars performed by Holmberg \& Flynn using the modified densities and dispersions -- but they are similar to those tabulated by Flynn et al. or Read and are good enough for our present purposes.

\begin{table}
\label{mass-model}
\caption{A local baryonic mass model using classic (updated) ISM density estimates.}
\begin{tabular}{lrrr}
\hline
Description & $\rho(0)^\dag$     & $\sigma_z^\dag$ & ${\Sigma_{HF}}^\star$ \\
            & ($\mMSUNCPC$) & ($\KPS$)   & ($\MSUNSPC$) \\
\hline
Molecular gas        & 21.0 (35.0)   &   4.0 &  3.1 (5.1) \\
Cold neutral medium  & 16.0 (32.0)   &   6.1 &  3.6 (7.3) \\
Warm neutral medium  & 12.0          &  14.8 &  7.3 \\
Warmer gas           &  0.9          &  40.0 &  2.0 \\
\hline
$M_V < 2.5$          &  3.0          &   7.5 &  0.8 \\
$2.5 < M_V < 3$      &  1.5          &  10.5 &  0.6 \\
$3 < M_V < 4$        &  1.9          &  14.0 &  1.1 \\
$4 < M_V < 5$        &  2.1          &  18.0 &  1.6 \\
$5 < M_V < 8$        &  6.8          &  18.5 &  5.4 \\
$M_V > 8$            & 13.1          &  18.5 & 10.5 \\
Giants               &  0.6          &  20.0 &  0.5 \\
White dwarfs         &  5.8          &  20.0 &  5.1 \\
Brown dwarfs         &  1.9          &  20.0 &  1.7 \\
\hline
Thick disc       &  4.4          &  37.0 &  8.6 \\
Stellar halo         &  0.1          & 100.0 &  0.7 \\
\hline
Sum of gas           & $50 (80)\pm 25$       &  -    & 16 (22) \\
Sum of stars         & $41\pm 4$     & -     & 37 \\
Total sum            & $91 (121)\pm 25$  & -     & 53 (58) \\
\hline
\end{tabular}\\
$^\dag$ Based on Flynn et al.\,(\CITE{2006MNRAS.372.1149F}); see text for details.\\
$^\star${Calculated assuming isothermal components and the potential of Holmberg \& Flynn (\CITE{2004MNRAS.352..440H}).}
\end{table}

The traditional estimates of the neutral and molecular gaseous densities from the 1980s and 1990s are off for a variety of reasons;
we now know that the density and ionization structure of the real interstellar medium (ISM) is much more complex. 
For example, a significant fraction of the HI cold neutral medium (CNM) is not optically thin, increasing the true densities by at least $\sim$30\% (Grenier, Casandjian \& Terrier \CITE{2005Sci...307.1292G}) and probably by as much as factors of 2-3 (Braun \CITE{2012ApJ...749...87B}; Fukui et al. \CITE{2014ApJ...796...59F}).
There is ``dark'' molecular gas not easily seen in CO but visible in gamma rays (e.g. Ackermann et al. \CITE{2012ApJ...755...22A}) and [CII] (e.g. Pineda et al. \CITE{2013A&A...554A.103P}) representing as much as 40\% of the total, i.e. an increase from the traditional value of 67\%.
These effects must raise the estimates for the local ISM densities considerably: assuming a factor of 2 correction for the CNM and a factor of 1.67 for the molecular gas, $\rho_{gas,local}(0)$ increases from $50$ to $80\,\mMSUNCPC$.
The Sun also sits near the middle of the Local Bubble (e.g. Lallemont et al. \CITE{2003A&A...411..447L}), a significant hole in the local Galactic ISM with a size comparable to the local thin-disc scale-height; the local gas density on somewhat larger scales is much higher by definition.

Unfortunately, the derived local value of $\rho_b(0)$ is not even what we really need: $\rho_{b,dyn}(0)$ and $\rho_{b,local}(0)$ are not the same, given that the potential of the MW disc experienced by the stars is highly non-uniform and that the vertical epicylic frequencies are $\sim$2-3 times larger than the orbital frequency (e.g. BT08, p. 167), effectively averaging out the local potential differences.
The Sun sits in a trough between two major spiral arms (Perseus and Scutum-Centaurus) and at the edge of the Orion Spur (Xu et al. \CITE{2013ApJ...769...15X}): the local volume- and surface-densities must then be uncharacteristically lower than the mean kinematical values -- again by definition -- unless the effects of the Orion Spur are uncharacteristically large.
It is extremely difficult to measure empirically or estimate theoretically how big this effect is, so it is generally not taken into account in the literature, not even  when fitting local or global DM models.
Drimmel \& Spergel (\CITE{2001ApJ...556..181D}) tried to measure the spiral amplitude of the MW by modelling projected COBE NIR data: they placed the Sun in an inter-arm region with a surface density contrast $\Sigma_{interarm}/\Sigma_{arm}\approx 0.85$ in the K-band, corresponding to a spiral amplitude $A=0.14$.
External galaxies may give us a more reliable estimate for the typical arm-interarm contrast of large spiral galaxies like the MW (Rix \& Zaritsky \CITE{1994ASSL..190..151R}).
Grosb{\o}l, Patsis \& Pompei (\CITE{2004A&A...423..849G}) analysed ground-based K-band images of 54 normal spiral galaxies and found amplitudes between 0.12 and 0.23.
Elmegreen et al. (\CITE{2011ApJ...737...32E}) and Kendall, Clarke \& Kennicutt (\CITE{2014arXiv1411.5792K})
studied 62 MW-like galaxies at $3.6\,\MU$ using the Spitzer satellite (Sheth et al. \CITE{2010PASP..122.1397S}; Kennicutt et al. \CITE{2003PASP..115..928K}) 
and found  typically larger ratios of $A=0.1-0.5$.
Thus, a galaxy like the MW is likely to have $A\!\approx\!0.2$, representing an interarm-to-mean surface density correction of $1/(1-A)\!\approx\!1.25$.
Vertical compression make the effects in {\it volume} density even larger than the ones in {\it surface} density: Dremmel \& Spergel (\CITE{2001ApJ...556..181D}) invoke a mid-plane arm-interarm flux density contrast that is $1.3$ times larger than the projected value, yielding an interarm-to-mean volume density correction of $1/(1-1.3A)\!\approx\!1.35$.

The effects of spiral arms are even stronger in the ISM: while Holwerda et al. (\CITE{2005A&A...444..109H}) find that there is only a difference of 2 in the opacities of arm and inter-arm regions, corresponding to the $A\approx 0.2$ seen in external stellar discs, they attribute this effect to cloud clumpiness and not to total surface density.
Indeed, the HI spiral amplitudes seen in the THINGS datasets
(Walter et al. \CITE{2008AJ....136.2563W}) 
are much larger than those seen for stars, reaching up to nearly 100\% amplitudes.
Thus, it is not surprising that the mean HI surface densities at the Galactic radius of the Sun estimated by Kalberla \& Dedes (\CITE{2008A&A...487..951K}) are larger than those of the local estimates\footnote{Read (\CITE{2014JPhG...41f3101R}) forgot to correct the Kalberla \& Dedes value of $\sim\!12\,\MSUNSPC$ from the pure hydrogen to the total density.}: $\overline{\Sigma(HI\!+\!He)} \approx 15\,\MSUNSPC$ (uncorrected for optical depth effects), consistent with a spiral amplitude $A \approx (15\!-\!11)/15 = 0.3$.

Assuming that $\rho_{dyn}(0) \approx \overline{\rho(0)} \approx \rho_{local}(0)/(1-1.3A)$, a spiral amplitude of 20\% alone increases the effective value of $\rho_{b,dyn}(0)$ from $91$ to $123\,\mMSUNCPC$ and with more modern estimates of the densities to $163\,\mMSUNCPC$.
In order to compare these values of $\rho_{dyn}(0)$ with those derived from recent kinematical studies (e.g. the exhaustive compilation in Read \CITE{2014JPhG...41f3101R}), one can calculate the effective value of $\rho_{DM,eff} \equiv \rho_{dyn}(0)-\rho_{b,min}(0)$ assuming Read's baryonic minimum estimate of $91\,\mMSUNCPC$: either the ISM and spiral amplitude corrections applied alone correspond to an apparent DM density of $\sim 30\,\mMSUNCPC$, and both corrections applied simultaneously result in $\sim 72\,\mMSUNCPC$.
Assuming that the estimates for $\rho_{dyn}(0)$ are accurate, it would appear that any reasonable correction to $\rho_b(0)$ for known effects is much more than is needed and corresponds to apparent DM densities that are much larger than the range $5-15\,\mMSUNCPC$ usually derived and thought to be consistent with cosmological expectations and galaxy-evolution scenarios.
This is already true for the traditional estimate of the local baryonic density, given the uncertainties in the traditional estimate of the gas contribution, and the realistic corrections to the total ISM densities and/or corrections for spiral structure only make the situation much worse.
Thus, practically {\it any} empirical value of $\rho_{dyn}(0)$ in the range found by current kinematic studies is incapable of constraining the local DM density without additional (presently unavailable) information, assumptions, and/or additional external constraints.


\section{Inhomogenous data and the effects of model assumptions}

\label{sec:inhomogeneous}

Since local estimates of $\rho_{DM}$ using current data are unable to yield robust values by themselves, we must turn to kinematical effects farther away from the plane of the MW (e.g. Bahcall \CITE{1984ApJ...287..926B}; Garbari et al.\,\CITE{2011MNRAS.416.2318G}).
By far the simplest method is to use the Jeans equations (moments of the Boltzmann equation; e.g. Sect. 4.8 in BT08), which connect the observed dispersions -- particularly the z-z component $\sigma_{zz}(z) \equiv \sigma_z(z)^2$ -- with the local gravity above and below the plane of the Galaxy.
The latter is dominated by the relatively well-constrained stellar populations, small amounts of gas, and (presumedly) a roughly constant DM density.
Here, there are also various approaches using different assumptions and amounts of data (e.g. 
Kuijken \& Gilmore \CITE{1989MNRAS.239..605K};
Bovy \& Tremaine \CITE{2012ApJ...756...89B}, hereafter BT12).

Moni Bidin et al. (\CITE{2012ApJ...751...30M}) analysed the kinematic data for a sample of colour-selected giants taken from 
Moni Bidin, Carraro \& Mendez (\CITE{2012ApJ...747..101M};
hereafter MBCM), for which the complete dispersion tensor could be estimated, using the Jeans and Poisson equations to derive the surface density with height, $\Sigma(z)$.
Assuming a double exponential tracer density distribution and a mean azimuthal velocity that is not a function of Galactocentric distance at any height from the plane,
they derived a local DM density from the asymptotic behaviour of $\Sigma(z)$  of $0\pm 1\,\mMSUNCPC$.
Bovy \& Tremaine (\CITE{2012ApJ...756...89B}) argued that one should assume a particular form for the vertical variation in the radial force field and a different radial scale-length, and so derived a value of $8\pm 3\,\mMSUNCPC$ from a simple vertical Jeans equation
(but see Moni Bidin et al. \CITE{2015A&A...573A..91M}).
Beyond difficult technical questions about the measurability of dynamical quantities and their gradients and -- in the case of the full Jeans equation -- the values of various global MW parameters, all of these analyses suffer from a major defect: they incorrectly assume that the MBCM data represent the kinematics and density of a uniform tracer population with a well-defined uniform scale-length and scale-height.
In fact, the giants are a vertically differentiated mixture of populations, each with its own very different abundance, scale-length, scale-height, and dispersion, and selection by colour alone is not enough to insure homogeneity of the spatially resolved kinematics.
Since the vertical dispersion is a strong function of scale-height, the varying mixture of tracer populations automatically creates a large positive gradient in the mean $\sigma_z$ as well as a decrease in the effective scale-height over the thick-disc value.
Because the Jeans equations are linear, spatial and kinematic inhomogeneity is not {\it \emph{per se}} a problem as long as the additional gradient in kinematics created by the mixture of tracer populations is accompanied by an equivalent change in the combined density.
Thus, the impact of kinematic inhomogeneity depends upon the model used.

Equation \,25 in BT12 -- derived assuming strictly exponential density distributions with scale-heights $h$ -- can be used to understand the effects in the MBCM data at heights $|z| >> h$: assuming the change in the derived value of $\Sigma(z)$ is mainly due to DM, that the effects of cross-dispersion terms are small, and that -- consistent with the current quality of the kinematic data -- any gradient in $\sigma_z^2(z)$ is roughly constant, one can see that the DM reveals itself mostly via an induced vertical gradient in $\sigma_z^2(z)$ at large heights:\\[-2ex]
\begin{equation}
\label{rhodm}
\rho_{DM} \approx \frac{1}{2} \left(\frac{\partial \Sigma(z)}{\partial z}\right)_{|z|>>h}  \approx \frac{1}{4\pi G h} \left(\frac{\partial \sigma_z^2}{\partial z} \right)_{|z|>>h}
.\end{equation}
However, if this equation is applied to an inhomogeneous mixture of tracer populations rather than a homogeneous thick-disc tracer, an increase in the gradient of $\sigma_z(z)$ appears due to the addition of tracers with lower $\sigma_z$ and smaller $h$.
If the density of this mixture is fit with a single exponential vertical density profile, the result will be a shorter apparent scale-height and the fitted DM density will be higher  because of the increase in the apparent dispersion gradient and because of the decrease in apparent scale-height.
%
%
%
One can easily estimate the magnitude of this effect using the simple thin+thick disc photometric model of Juri\'c et al. (\CITE{2008ApJ...673..864J}):
for two exponential components with scale-heights of 300 and 900\,pc, a local thick-to-thin density ratio of 0.12, reasonable isothermal dispersions of 15 and $45\,\KPS$ (Kordopatis et al. 2013), and an observed value simply weighted by the relative densities, the tracer mixture effect alone is expected to produce a 20\% increase in $\sigma_z$ over the range of 1.4 to $4.5\,\KPC$ covered by the MBCM data, exactly as  is observed.
The corresponding change in effective scale-height depends upon the details of the sampling and fitting, but simple mock datasets created using the above model show a significant decrease in the effective vertical scale-height from the thick-disc value assumed by MB12.
Bovy \& Tremaine (\CITE{2012ApJ...756...89B}) attempted to correct for the uncertainty in the mean scale-height, but if an inhomogeneous distribution is non-exponential, the fundamental equation and not the parameter is the problem.
Sanders (\CITE{2012MNRAS.425.2228S}) has performed a much more detailed simulation of the MBCM dataset and analysis using DF, a simple thin and thick disc, an assumed evolutionary history, a larger (20\%) local thick disc fraction, and a standard DM halo.
He -- of course -- found the same increase in the gradient in $\sigma_z^2(z)$, but the complexity of his model has led to the impression that the increased gradient implies {\it \emph{more}} rather than {\it \emph{less}} DM (Garbari et al. \CITE{2012MNRAS.425.1445G}; Piffl et al. \CITE{2014MNRAS.445.3133P}). 

A  similar effect plagues the analysis of Bienaym\'e et al. (\CITE{2014A&A...571A..92B}), who used RAVE observations of red clump giants, a simple potential model, and an assumed DF to derive the vertical gravity $g_z(z)$.
Fitting $g_z(z)$ {\it \emph{a posterori}} with a  single doubly exponential disc, an intermediate radial scale-length of $2.5\,\KPC,$ and a constant DM density, they then found $\rho_{DM}=14\pm 1\,\mMSUNCPC$.
Bienaym\'e et al.\,did split their kinematic dataset into three abundance groups, corresponding to [Fe/H] $<$ -0.58, [Fe/H] $>$ -0.25, and an intermediate population; while not enough to isolate homogenous tracers, this should result in similar vertical scale-heights (see Fig.\,4 in Bovy, Rix, Liu et al. \CITE{2012ApJ...753..148B}), so that the density-weighted $\sigma_z$ should not be a strong function of height even if each subsample is kinematically inhomogenous.
However, interstellar extinction makes the radial range over which these samples were obtained a strong function of galactic latitude: the thin-disc data was obtained over a significantly smaller range of Galactic radii than the thick-disc data.
Chen et al. (\CITE{2012ApJ...752...51C}) showed that the thin- and thick-disc scale-lengths estimated from SEGUE stars near the Galactic plane are different, confirming the results of Bensby et al. (\CITE{2011ApJ...735L..46B}) using many fewer stars: their values are $H_{thin}\!=\!3.4$ and $H_{thick}\!=\!1.8\,\KPC$, suggesting that the thin-disc scale-length is nearly twice that of the thick disc.
Assuming that the squared velocity dispersions vary exponentially with the same scale-length and hence are larger at smaller Galactic radii, the thick-disc data is weighted to smaller average radii and hence larger dispersions.
Thus, there must be an inverse correlation between the radial scale-length and $\sigma_z$, producing an artificially larger dispersion and hence larger $\rho_{DM}$ for the data thought to be the most sensitive.
Again, a crude model can be used to estimate the effects: for an object selection criterion roughly corresponding to that used by Bienaym\'e et al., consisting of a minimum distance of $200\,\PC$, a maximum distance of $3000\,\PC$, $|b|\!>\!22^\circ$, and radial scale-lengths of $1.6$ or $3.4\,\KPC$ for the thick and thin disc, respectively, one expects an artificial increase in the relative vertically averaged values of the thick-disc $\sigma_z$ over those of the thin disc over the range of 200 to $2000\,\PC$ of $\sim 20$\% independent of the common vertical scale-height.
This is roughly the gradient seen in the RAVE data and a sampling effect not considered in their model.

Even when considerable effort is put into an attempt to correct for systematic effects, to treat consistently inhomogenous tracers, and to minimized the number of assumptions, the result is not necessarily more reliable.
For instance, Garbari, Read \& Lake (\CITE{2011MNRAS.416.2318G}) use a detailed global MW model in an attempt to correct for the effects of a realistic Galaxy (e.g. spiral arms), but their mock-data simulation has neither a thick-disc  component  nor an ISM component even though these two components make up 60-70\% of the mid-plane density (Table\,1) and differences in radial scale-lengths between the thin and thick discs must also affect the sampling.
In addition, their fits were constrained to have $\rho_{b,local}(0)$ in the unnecessarily narrow range $91\pm 14\,\mMSUNCPC$, which implies a considerable shift in kinematic influence to the otherwise less well-constrained DM halo.

The  problems with mixed tracer populations can be minimized by carefully choosing each sample to be as homogenous as possible, and the most obvious criterion of homogeneity is detailed abundance, e.g. [Fe/H] and/or [$\alpha$/Fe].
Zhang et al. (\CITE{2013ApJ...772..108Z};
hereafter Z13) derived a DM density of $6.5\pm 2.3\,\mMSUNCPC$ using three different [$\alpha$/Fe]-selected, K-dwarf tracer populations taken from the SEGUE survey at vertical heights between 300 and 1400\,pc, roughly corresponding to thin-, intermediate-, and thick-disc tracer populations.
However, there are several problems with this analysis as well:
Z13 used an approximate vertical Jeans equation;
they assumed that the baryonic contributions to the local vertical gravitational field are due to infinite homogenous discs; 
their stellar disc had a thin-disc scale-height even though at the higher heights the gravity should be dominated by the thick disc;
they assumed the traditionally low gas surface density (see discussion in the previous section);
the three (assumed exponential) tracer populations each cover a wide range in [Fe/H] and hence in scale-heights, in all probability producing the kinematic inconsistencies implied by Eq.\,\ref{rhodm} (the thin and intermediate tracers have density profiles with non-exponential tails, as expected);
and they used an undocumented Markov chain Monte Carlo  (MCMC) prior on the scale-heights (Zhang, private communication) without which their value of $\rho_{DM}$ is undetermined.

Bovy \& Rix (\CITE{2013ApJ...779..115B}) analysed a SEGUE/SDSS dataset of G dwarfs at Galactic radii $4.5\!<\!R\!<\!7\,\KPC$ and much larger heights than was possible with the K dwarfs used by Z13.
They modelled the data using DF models for each mono-abundance ([Fe/H] and [$\alpha$/Fe]) data subset and modified St\"ackel potentials,
permitting them to simultaneously estimate important quantities like the radial scale-lengths.
Many publications now use their derived value of the stellar surface density as an assumed accurate measure of the total baryonic contribution (e.g. Iocco et al. \CITE{2015NatPh..11..245I}). 
However, the use of St\"ackel potentials results in vertical dispersion profiles $\sigma_z(z)$ that increase towards the disc mid-plane for heights below 2\,kpc, and are otherwise flat (their Fig.\,4d); measurements of the near-mid-plane dispersions are significantly lower (Kuijken \& Gilmore \CITE{1989MNRAS.239..651K}; Cr\'ez\'e et al. \CITE{1998A&A...329..920C}; Dehnen \& Binney \CITE{1998MNRAS.294..429D}; Holmberg, Nordstr\"om \& Andersen \CITE{2007A&A...475..519H}; Pasetto et al. \CITE{2012A&A...547A..71P}) and the mono-abundance data of B\"udenbender et al. (\CITE{2014arXiv1407.4808B}; hereafter B14) clearly show that the gradient at such heights, if present, has the opposite sign.
Part of this problem may be due to the fact that their model did not account for the effects of the local inter-arm region (see the discussion in the previous section), which would produce an apparent change in the radial scale-length and hence potentially observable changes in the modelled kinematics.
They also used the same traditionally low estimate for the gaseous surface density and a single-component disc potential.
Other issues are discussed by Read (\CITE{2014JPhG...41f3101R}).
Thus, while Bovy \& Rix (\CITE{2013ApJ...779..115B}) have shown how potentially powerful the complicated method for modelling the kinematics of disc stars can be, their impressively precise results are, in fact, inaccurate at a level that is important when trying to measure $\rho_{DM}$.

B\"udenbender et al. (\CITE{2014arXiv1407.4808B}) separated the SEGUE G dwarf data into clearly thin- and thick-disc tracers and showed that the simple model used by K13 is unable to simultaneously explain  the differences in the magnitudes and the slopes of the two tracers.
They attributed this to different orientations of the velocity ellipsoid with height, implying that the vertical motions are not fully decoupled from the radial ones.
However, both the thin- and thick-disc tracer populations were not mono-abundance selected and, like the Z13 samples, contained stars with very different [Fe/H] and hence vertical scale-heights.
Was the difference due to model assumptions and fundamental kinematic effects, or simply due to inhomogeneous tracers?


\begin{figure} 
\label{fig:glocal}
\begin{center}
\includegraphics[width=9.2cm,clip=true, trim=0 118 0 15]{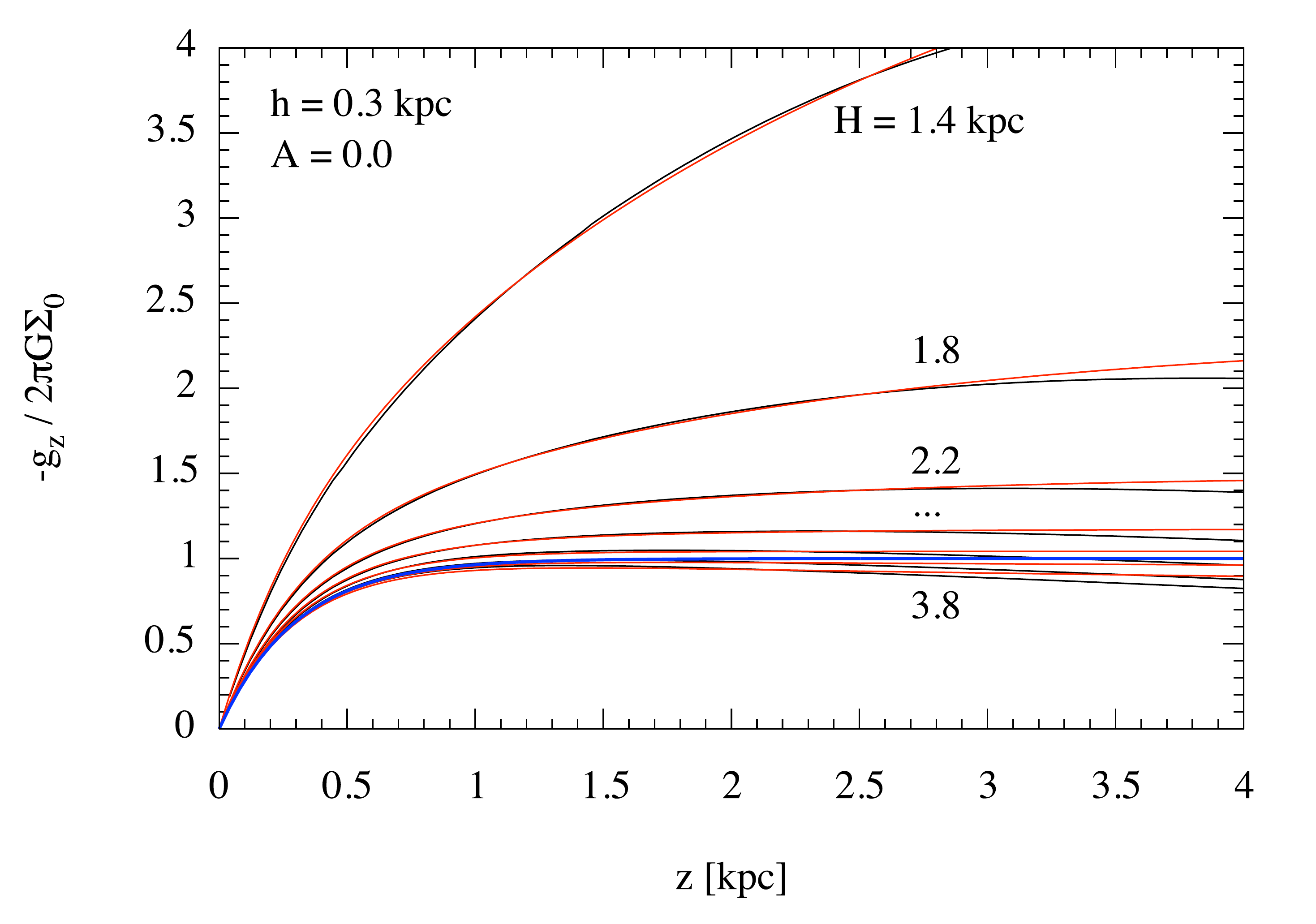}\\
\includegraphics[width=9.2cm,clip=true, trim=0 118 0 15]{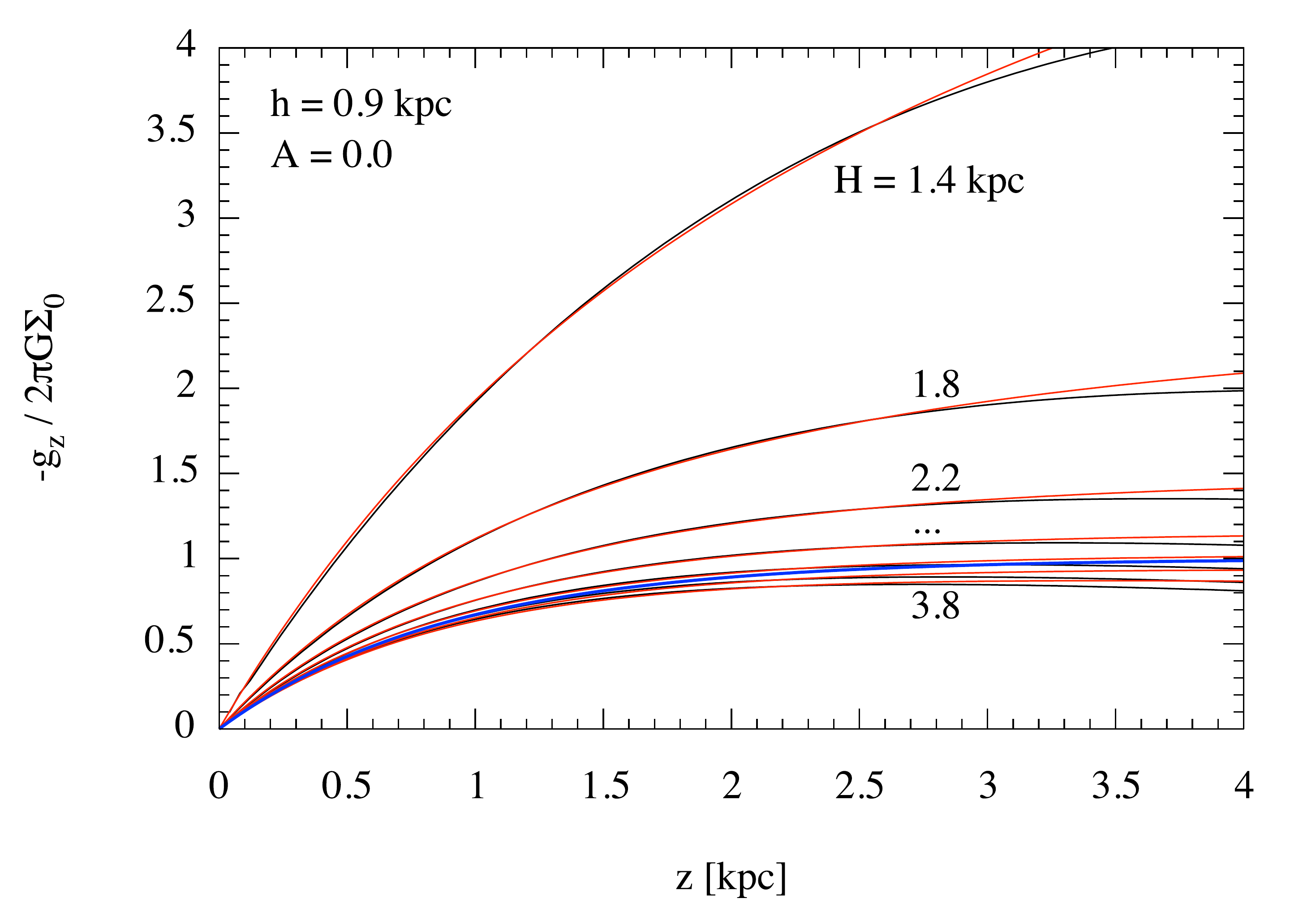}\\
\includegraphics[width=9.2cm,clip=true, trim=0  20 0 15]{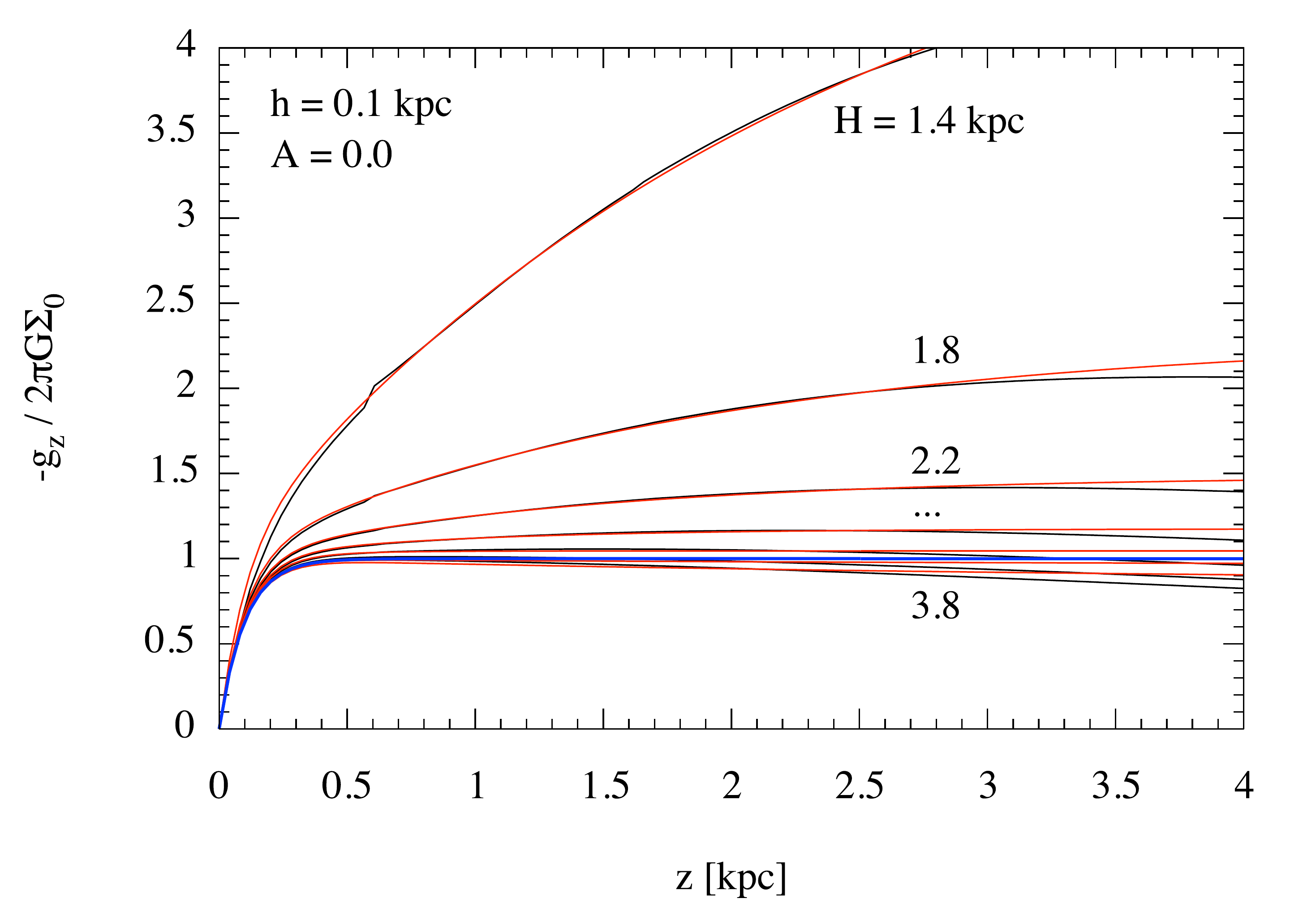}\\
\end{center}
\caption{
The vertical gravity $g_z$ of axisymmetric doubly exponential MW discs and radial scale-lengths $H=1.4, 1.8, 2.2, 2.6, 3.0, 3.4$, and $3.8\,\KPC$; the blue line is $g_z$ for a uniform infinite plane with the default local surface density $\Sigma_0\!\equiv\!\Sigma(R_0;A\!\equiv\!0)$; the red lines are fits to $g_z(z<3\,\KPC)$ assuming an additional effective infinite uniform disc component (see text).}
\end{figure}

\section{The not-so-local gravity field}

\label{sec:gravity}

In order to derive $\rho_{DM}$ from homogenous velocity dispersion data using a minimally complicated model of the MW -- stationary, axisymmetric, vertically symmetric -- in a manner similar to that used by Z13, one must specify the gravitational forces on the stars more accurately than possible in the simple model they used.
Although the difference between the gravity from an infinite, uniform disc and the real MW should be small for $|z|\!<<\!R_0$, the MBCM data reach to $\sim\!R_0/2$ and a correction to the model of an infinite uniform disc that takes structure at larger scales is needed.
Unfortunately, the effort to construct a global mass model is exactly what one wanted to avoid by solving the vertical Jeans equation alone, so the question is whether the non-uniform effects are significant and, if so, how easily they can be included as corrections to a local analysis.

The effects due to the global variation of the disc surface density with radius, parameterised by the radial scale-length $H$ of some component, are straightforward to calculate and are often -- but by no means always -- included in kinematical analyses when they should be.
For large $H$, the disc regions at smaller and larger Galactic radii will have similar surface densities and one would naively expect the correction to be small.
For small $H$, the surface density at smaller Galactic radii will have a significantly larger surface density that could dominate the local vertical gravity contribution and which may or may not be compensated by the loss of surface density at larger radii -- the gravitational field of a planar annulus is very different from that of a point source, making it hard to know the outcome {\it \emph{a priori}}.
For a doubly exponential disc component, this effect can be calculated semi-analytically using the expressions for $g_z(R,z)$
derived by Kuijken \& Gilmore (\CITE{1989MNRAS.239..571K})\footnote{The vertical acceleration $g_z$ is often labelled $K_z$; the latter often has an ill-defined sign, and is often incorrectly called a ``force'' in the literature.}.

The results for a doubly exponential disc with vertical scale-height $h=300\,\PC$ and a range of $H$ are shown in Fig.\,1 (top): for $H\!\approx\!3\,\KPC$, $g_z$ is very similar to that due to an infinite uniform disc; for larger $H$ there is a slight decrease at large $|z|\!>>\!h$; and
for smaller $H$ there is a substantial increase in magnitude and the lack of a flattening at large $|z|\!>>\!h$.
Note that the last effects increase the contribution to $g_z$ from the baryonic discs -- perhaps dramatically -- compared to a model that uses local disc properties only, and the second  one mimics the effects of a constant DM density ($g_z(disc) \not\approx const$ as $|z| \rightarrow \infty$); both must result in a reduction in the amount of DM needed to fit the data.
This is the opposite effect to that reported by Kuijken \& Gilmore (\CITE{1989MNRAS.239..651K}): in their Fig.\,2, they assumed a smaller $R_0 = 7.4\,\KPC$ and a larger $H = 4.5\,\KPC$, i.e. $H/R_0=0.6$.
For values of $H/R_0$ near unity, the gravitational field of the disc starts to approach that of a point mass, leading to the drop in $g_z(z)$ at large heights.

The change in the {\it form} of $g_z$ relative to that of an infinite disc is itself very similar to that of an infinite uniform disc with a different surface density and scale-height: in Fig.\,1, the fits to the real form of $g_z(z)$ consists of that for an infinite disc with the nominal local surface density plus an additional infinite disc component with fitted surface density and scale-height (e.g. the decrease in $g_z(z)$ at large $|z|$ for large $H$ is then modelled as a negative additional surface density).
When the Juri\'c et al. (\CITE{2008ApJ...673..864J}) thin- and thick-disc vertical scale-heights $h$ of 300 and $900\,\PC$ and the corresponding radial scale-lengths $H$ from Bensby et al. (\CITE{2011ApJ...735L..46B}) of 3.4 and $1.8\,\KPC$, respectively, are used, the gravity due to the thin disc is nearly unaffected whereas the effective thick-disc surface density increases by 140\%, resulting in an increase in the total effective surface density of 37\%, and the additional component has a large effective vertical scale-length of $2.6\,\KPC$.
%
%
The changes in the form of $g_z(z)$ for the very thin gaseous disc of the MW is negligible because the radial scale-length is roughly twice that of the disc (Bigiel \& Blitz \CITE{2012ApJ...756..183B}).

Thus, the effects on the vertical gravity profile of non-uniform stellar structures in the Galactic neighbourhood of the Sun are quite significant and must be included in all vertical Jeans analyses.
The naive separation of the gravity into an asymptotically constant disc and increasing DM component (e.g. the classic Kuijken \& Gilmore ``K'' and ``F'' models) is not necessarily an adequate representation.
Fortunately, even though the local vertical gravity is not that of an infinite disc with constant surface density, the effects of a realistic axisymmetric mass-distribution for the disc of the MW can be modelled using additional infinite disc components whose surface densities and scale-heights depend simply upon the true values.


\begin{figure*}
\begin{center}
\label{fig:sigma_R}
\includegraphics[width=12.0cm]{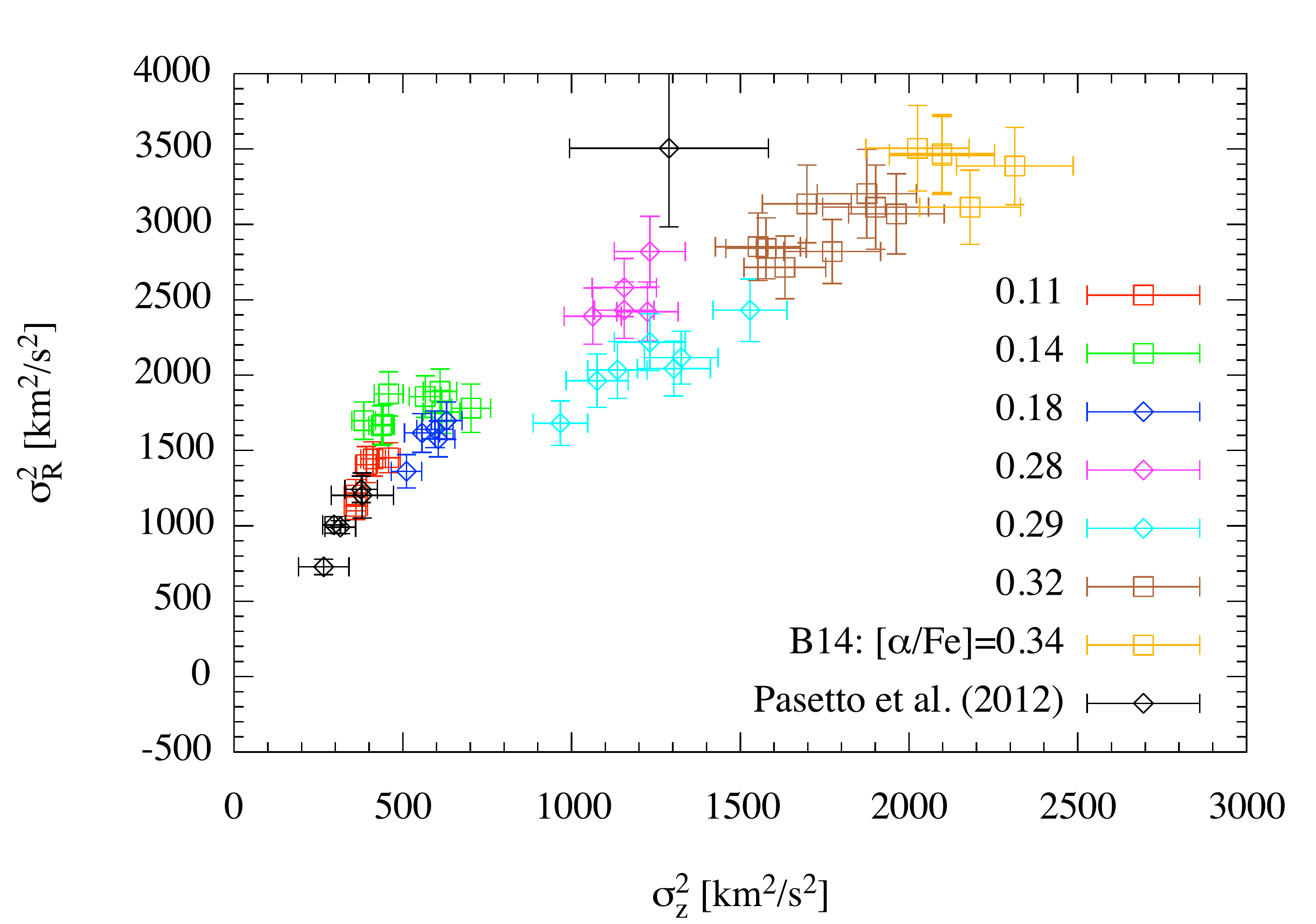} \\
\caption{ 
$\sigma_z^2$ and $\sigma_R^2$ data from B\"udenbender et al. (2014; B14) and Pasetto et al. (2012a,b); for the B14 SEGUE data, each colour represents a different [Fe/H] and [$\alpha$/Fe] mono-abundance tracer population (and was chosen to reproduce the corresponding plots in B14), whereas the Pasetto et al. data are for undifferentiated RAVE stars.
}
\end{center}
\end{figure*}

\section{The vertical Jeans equation revisited}

\label{sec:jeans}

The vertical Jeans equation can be easily derived from moments of the collisionless Boltzmann equation in cylindrical coordinates for a steady, axisymmetric stellar tracer component $i$ with density $\rho_i(R,z)$ (see BT08; Eq. 4.222b, p. 353).
For a steady-state situation ($\overline{V_k}=0$ for $k=R,z$, i.e. no net flux of stars in $R$ or $z$), one obtains the vertical, axisymmetric Jeans equation
%
\begin{equation}
\label{Jeans}
\frac{\partial}{\partial z} \left( \rho_i \sigma_{zz}^{(i)} \right) 
 + \frac{1}{R} \frac{\partial}{\partial R} \left( R \rho_i \sigma_{Rz}^{(i)} \right)
 - \rho_i g_z = 0
.\end{equation}
%
This is an exact and full description of the connection between the local vertical gravity and these local stellar velocity dispersions within the very modest assumptions made.
Admittedly, there is evidence that the vertical structure of the MW's thin disc is not stationary (e.g. 
Widrow et al. \CITE{2012ApJ...750L..41W}, Carlin et al. \CITE{2014ApJ...784L..46C}),
but these are second-order effects that should not seriously affect the velocity dispersions, particularly at larger heights.
Admittedly, the gravitational field of the MW is not axisymmetric owing to factors like the central bar and the spiral arms, but the effects of the former are subtle (e.g. 
Bovy et al. \CITE{2014ApJ...790..127B}) and the non-axisymmetric effects of the latter are  small due to local pitch angles of less than $20^\circ$ (Vall\'ee \CITE{2005AJ....130..569V}; Levine, Blitz \& Heiles \CITE{2006Sci...312.1773L}).

Fortunately, the mean radial surface brightnesses of disc stars in most spiral galaxies are known to be exponential
(Freeman \CITE{1970ApJ...160..811F}) and the local vertical distributions of stars in the local MW are also observed to be exponential starting at quite modest heights above the plane of the disc
(Juri\'c et al. \CITE{2008ApJ...673..864J}).
In particular, this holds for mono-abundance tracers
(Bovy, Rix \& Hogg \CITE{2012ApJ...751..131B};
Bovy et al. \CITE{2012ApJ...753..148B};
K13) so that one can confidently assume a doubly exponential density profile for any homogenous tracer population, and simply measure the radial scale-lengths $H_i$ and vertical scale-heights $h_i$ in order to characterise this most important property to an adequate accuracy :
\begin{equation}
\rho_i(R,z) \approx \rho_i(R_0,0) e^{-(R-R_0)/H_i} e^{-|z|/h_i}
.\end{equation}
Given the observed properties of $\rho_i$, it is not unreasonable to make a similar assumption about the radial variation of the velocity dispersion:
\begin{equation}
\label{L}
\sigma_{zz}^{(i)}  \propto e^{-(R-R_0)/L_z^{(i)}}
.\end{equation}
A simple exponential disc with a constant disc thickness $h$ requires $\sigma_{zz} \propto \Sigma$, i.e. $L_z \approx H$.
More complicated dynamical models of the MW suggest that $L_z$ is larger than $H$ by $\sim$20\% (e.g. Sanders \& Binney \CITE{2015arXiv150102227S}), so assuming they are equal should not be a major source of error.

To solve the vertical Jeans equation fully, one needs to know the radial and vertical behaviour of $\sigma_{Rz}$.
This component generally has a smaller effect than $\sigma_{zz}$ because the $h_i$ are generally much smaller than the $H_i$, so this contribution was totally left out in Z13.
Several possible models for $\sigma_{Rz}$ have  been proposed (see Amendt \& Cuddeford \CITE{1991ApJ...368...79A}).
For example, one can assume that the principle axes of the velocity ellipsoid are aligned in spherical or cylindrical coordinates so that the correlation between the radial and vertical motions can be expressed via
a tilt angle $\alpha_{tilt}$ such that
\begin{equation}
\label{tilt-angle}
\sigma_{Rz} 
\approx \frac{1}{2} \tan(2\alpha_{tilt}) \left(\sigma_R^2-\sigma_z^2\right)
\end{equation}
(see BT08, problem 4.32, p.\,391), where $\sigma_R^2\!\equiv\!\sigma_{RR}$, $\sigma_z^2\!\equiv\!\sigma_{zz}$.
For the special case of St\"ackel potentials, it can be shown that $\alpha_{tilt}$ is aligned with the potential gradients and so should be independent of the tracer population used (e.g. Statler \CITE{1989ApJ...344..217S}, Binney \CITE{2012MNRAS.426.1324B}).
This form was used by
Siebert et al. (\CITE{2008MNRAS.391..793S}) and
Cassetti-Dinescu et al. (\CITE{2011ApJ...728....7C})
to analyse abundance-undifferentiated data from the (southern) RAVE experiment close to the Galactic plane.
B\"udenbender et al. (\CITE{2014arXiv1407.4808B}) have measured $\alpha_{tilt}$ for the abundance-differentiated dwarf G stars in the (northern) SEGUE sample and saw that $\alpha_{tilt}$ increases in magnitude with increasing $z$: they fitted $\alpha_{tilt}$ as a linear function in $z$
but, given the errors, an equally good fit to the data is simply
\begin{equation}
\label{tan2alpha}
\tan(2\alpha_{tilt}(z)) ~ = ~ (-1.88\pm 0.19)\frac{z}{R_0} ~~ \equiv ~~ c \frac{z}{R_0}. \\ \nonumber
\end{equation}
Both fits imply that the tilt angle is essentially aligned with the local spherical coordinate system.
If an anisotropy parameter $\beta(R,z) \equiv \sigma_{z}^2/\sigma_{R}^2$ is defined (e.g. Pasetto et al. \CITE{2012A&A...547A..70P}), the assumptions $\beta\equiv const$ plus a tilt angle perfectly aligned with the Galactic center result in
\begin{equation}
\label{KuijkenGilmore}
\sigma_{Rz} \approx \left(\frac{1}{1+\frac{z^2}{\beta R^2}}\right) \frac{z}{R} \left(\sigma_R^2-\sigma_z^2\right) 
\end{equation}
(Kuijken \& Gilmore \CITE{1989MNRAS.239..571K}), implying that the assumption $c \approx const$ is not unreasonable for $z^2 << \beta R^2$.

The SEGUE data from the  B14 Table\,1  are shown in Fig.\,2, covering heights $0.5 < z< 2.5\,\KPC$.
Clearly, there is a tight connection between  $\sigma_R^2$ and $\sigma_z^2$ that appears to be quite linear within a given tracer population and very similar for different tracer populations.\footnote{Actually, all of the data are  consistent with the single assumption $\sigma_R^2\!=\!(68.73\pm 1.1\,\KPS)\sigma_z$, but there is no obvious reason for this relation and it is not very useful for the analysis that follows.}
If one generically assumes
\begin{equation}
\label{sigmaR}
\sigma_{RR}^{(i)}(R,z) \approx a_i + b_i \sigma_{zz}^{(i)}(R,z)
\end{equation}
(i.e. $\beta_i\equiv 1/b_i$ for $a_i\equiv 0$),
one obtains a single equation expressing $\sigma_{Rz}$ in terms of $\sigma_{z}$, $z$, and two dispersion-model parameters $\eta$ and $\gamma$,
%
\begin{eqnarray}
\label{sigmaRz}
\sigma_{Rz}^{(i)}(R,z)
        & \approx & 
        \frac{1}{2} c_i \frac{z}{R} \left( a_i + (b_i-1) \sigma_{zz}^{(i)}(R,z) \right) 
        \\ \nonumber
        & \equiv &
        -\left(\eta_i + \gamma_i \sigma_{zz}^{(i)}(R,z) \right) \frac{z}{R}
\end{eqnarray}
%
where $\eta \equiv -ca/2$ and $\gamma \equiv -c(b-1)/2$ (the sign was chosen to fit the northern SEGUE data of B14).
While this expression was derived using the tilt angle formalism, it constitutes a simple, phenomenological model for $\sigma_{Rz}$ in terms of the model parameters $\gamma$ and $\eta$ and an assumed scaling in $z$ that could cover other origins for the behaviour  of this dispersion component.
The relatively simple assumption that $\sigma_{Rz}$ is proportional to a linear function in $\sigma_{zz}$ whose constant term might be non-zero should not be construed as a potential physical problem: the fact that the function is non-zero for $\sigma_{zz}\!=\!0$ is irrelevant given that the dispersions are nearly isothermal for any given tracer population, and that what is needed is merely a functional connection between the two dispersions measures.

The usefulness of this model for $\sigma_{Rz}$ depends, of course, upon how well it fits the data, at least locally.
In order to augment the B14 data with observations at lower heights, the thin-disc RAVE results from Pasetto et al. (\CITE{2012A&A...547A..71P}) were modelled using Eq.\,\ref{L} for $\sigma_{zz}$ and the epicyclic approximation for $\sigma_{RR}$
(Amendt \& Cuddeford \CITE{1991ApJ...368...79A}).
The fits to $\sigma_{kk}(R_0,z)$ were made using the python MCMC object {\tt emcee} (Foreman-Mackey et al. \CITE{2013PASP..125..306F}) with the parameters $L_k$ and the $\sigma_k$ values at each height bin (0, $\pm$200, and $\pm 400\,\PC$).
A standard probability function $\ln P\!\propto\!-\chi^2$ and 30 MCMC walkers were used; after a burn-in of 2000 iterations each, the walkers were well converged and the probability density functions (PDF) were calculated from a total of 150000 samples.
While the values of $\sigma_{zz}(R_0,z)$ and $\sigma_{RR}(R_0,z)$ are reasonably well constrained, the final PDFs of the scale-lengths have long tails; if the radial scale-lengths are restricted to be with the reasonable range $0\!<\!L_k\!<\!10\,\KPC$, $L_z = 1.9\PM{\infty}{0.1}\,\KPC$, and $L_R = 2.8\,\PM{\infty}{0.1}\,\KPC$ (68.2\% bounds), consistent with being equal to a thin-disc density scale-length of $\sim\!3\,\KPC$.
When the RAVE dispersion points are added to Fig.\,2, it appears that $\eta\equiv 0$, ~ $\beta\!\approx\!const$ is a reasonable assumption for the thin disc, but the B14 thick-disc data shows a much flatter relation.
Thus, Eq.\,\ref{sigmaRz} appears to be an adequate -- if not excellent -- local phenomenological model that is able to model the potentially different behaviours of different tracers.

With these assumptions, one can then rewrite the local vertical Jeans equation as the ordinary differential equation
%
%
\begin{equation}
\label{finalJeans}
\frac{d\sigma_{zz}^{(i)}(R_0,z)}{dz} \!\!
        \approx
        g_z(R_0,z) +
        \left[
                \frac{1}{h_i} \frac{z}{|z|}
                -
                \frac{z}{2\lambda_i^2}
        \right] \sigma_{zz}^{(i)}(R_0,z)
        - \frac{\eta_i}{H_i} \frac{z}{R_0}
,\end{equation}
where
\begin{eqnarray}
\label{lambda}
\lambda_i
        & \equiv & \!\!
        \sqrt{\frac{R_0}{2 \left(\frac{1}{H_i}+\frac{1}{L_i}\right) \gamma_i}}
        ~ \approx ~ \sqrt{\frac{R_0 H_i}{4\gamma_i}}
        \\ \nonumber
        & \approx &
        2.2\,\KPC \left(\frac{H_i}{2.5\,\KPC}\right)^{1/2} \! \left(\frac{R_0}{8.2\,\KPC}\right)^{1/2} \gamma_i^{-1/2}
\end{eqnarray}
%
is a typical scale-length where the effects of cross-talk between $\sigma_{Rz}$ and $\sigma_{zz}$ appear, and $\lambda_i/h_i$ is a measure of the effects of the $\sigma_{Rz}$ term on $\sigma_{zz}$ for a given tracer: larger values imply less effect.
The $\eta_i$ term in Eq.\,\ref{finalJeans} is {\it not} proportional to $\sigma_{zz}$ and so behaves like a gravity term.
Indeed, the proportionality to $z$ makes it look like a constant apparent mass density $\rho_\eta^{(i)}$ defined by
\begin{eqnarray}
\label{rhoeta}
\rho_\eta^{(i)} \!
        & \!\equiv\!  & \!\!
        \frac{\eta_i}{4\pi G R_0 H_i}
        \\ \nonumber
        & \approx & \!\!
        1.5\,\mMSUNCPC ~ \left( \frac{\eta_i}{2000\,\KMS2} \right) \left( \frac{R_0}{8.2\,\KPC} \right)^{-1} \left( \frac{H_i}{3\,\KPC} \right)^{-1}, 
\end{eqnarray}
%
which represents a small but tracer-dependent and hence potentially noticeable correction to any fitted dark matter density for reasonable values of $\eta_i$ and $H_i$.

The equation for $g_z(R_0,z)$ used by Z13 to close the equation assumes an infinite uniform disc plus a constant DM density.
In Section\,\ref{sec:gravity}, we saw that a generalisation of their model -- the sum of effective infinite discs with different effective surface densities $\Sigma_j$ and effective exponential vertical scale-heights $l_j$ (the latter chosen to be easily distinguishable from the tracer scale-heights $h_i$), plus a constant DM density $\rho_{DM}(R_0,0)$ -- is a reasonable functional model:
\begin{equation}
\label{gravity}
\frac{g_z(R_0,z)}{2\pi G} \approx
        - 2\rho_{DM}(R_0,0) z
        - \sum_{j=1}^{M} \Sigma_j(R_0) (1-e^{-|z|/l_j}) \frac{z}{|z|} 
.\end{equation}
This assumption also permits one to solve the vertical Jeans equation (Eq.\,\ref{finalJeans}) analytically when the simple model for $\sigma_{Rz}$ (Eqs.\,\ref{sigmaRz} and \ref{rhoeta}) is used,
%
\begin{equation}
\label{solution}
\frac{\sigma_{zz}^{(i)}(z)}{2\pi G h_i}
        \approx
        2 \left(\rho_{DM}+\rho_\eta^{(i)}\right) h_i \Psi(z;h_i,\lambda_i) 
        + \sum_{j=1}^{M} \Sigma_j \Upsilon(z;h_i,\lambda_i,l_j)
,\end{equation}
where
\begin{eqnarray}
\Psi(z;h_i,\lambda_i)
        & \equiv &
        4 \frac{\lambda_i^3}{h_i^3} D\left(\frac{\lambda_i}{h_i}-\frac{|z|}{2\lambda_i}\right) - 2 \frac{\lambda_i^2}{h_i^2}
        \\ \nonumber
\Upsilon(z;h_i,\lambda_i,l_j)
        & \equiv &
        2 \frac{\lambda_i}{h_i} \left[ D\left(\frac{\lambda_i}{h_i}\!-\!\frac{|z|}{2\lambda_i}\right)
        - D\left(\frac{\lambda_i}{h_i}\!+\!\frac{\lambda_i}{l_j}\!-\!\frac{|z|}{2\lambda_i}\right) e^{-|z|/l_j} \right]
\end{eqnarray}
%
and where
\begin{equation}
D(x) \equiv e^{-x^2} \int_0^{x} e^{t^2} dt = \frac{\sqrt{\pi}}{2} e^{-x^2} \ERFI(x)
\end{equation}
is Dawson's function (erfi(x) is the imaginary error function).
This solution is only good for reasonable heights $<<\!2\lambda_i^2/h_i$, but this is not a critical constraint on the applicability of the solution since $\lambda_i/h_i\!>>\!1$ for any reasonable value of $\gamma_i$ and $h_i$ (Eq.\,\ref{finalJeans}) and the whole procedure breaks down (e.g. the assumptions of exponential density profiles and constant DM densities) if carried out too far from the disc plane.

The behaviour of this solution relative to that of Z13 is governed by the ratio $\lambda_i/h_i$, as is shown in Fig.\,3.
The effects of $\lambda_i$ are just apparent for $\lambda_i/h_i\!\sim\!16,$ but are quite significant for smaller values.
Thus, $\gamma_i$ can significantly change the form of the solution.
In the limit $\lambda_i\!\rightarrow\!\infty$, i.e. $\gamma_i\!\rightarrow\!0$, this solution is the same as that used by Z13 and B14.
However, for small $\lambda_i/h_i$, the simple asymptotic behaviour of the Z13 solution (flat disc and linear DM contributions to $\sigma_{z}^2$) is modified so that {\it \emph{both}} mass components produce increasing $\sigma_{z}$ with height, making both the baryons and the DM more effective at explaining positive dispersion gradients and thus principally reducing the total amount of DM needed.

\begin{figure}
\label{fig:functions}
\begin{center}
\includegraphics[width=9.2cm,clip=true,trim=30 112 0 10]{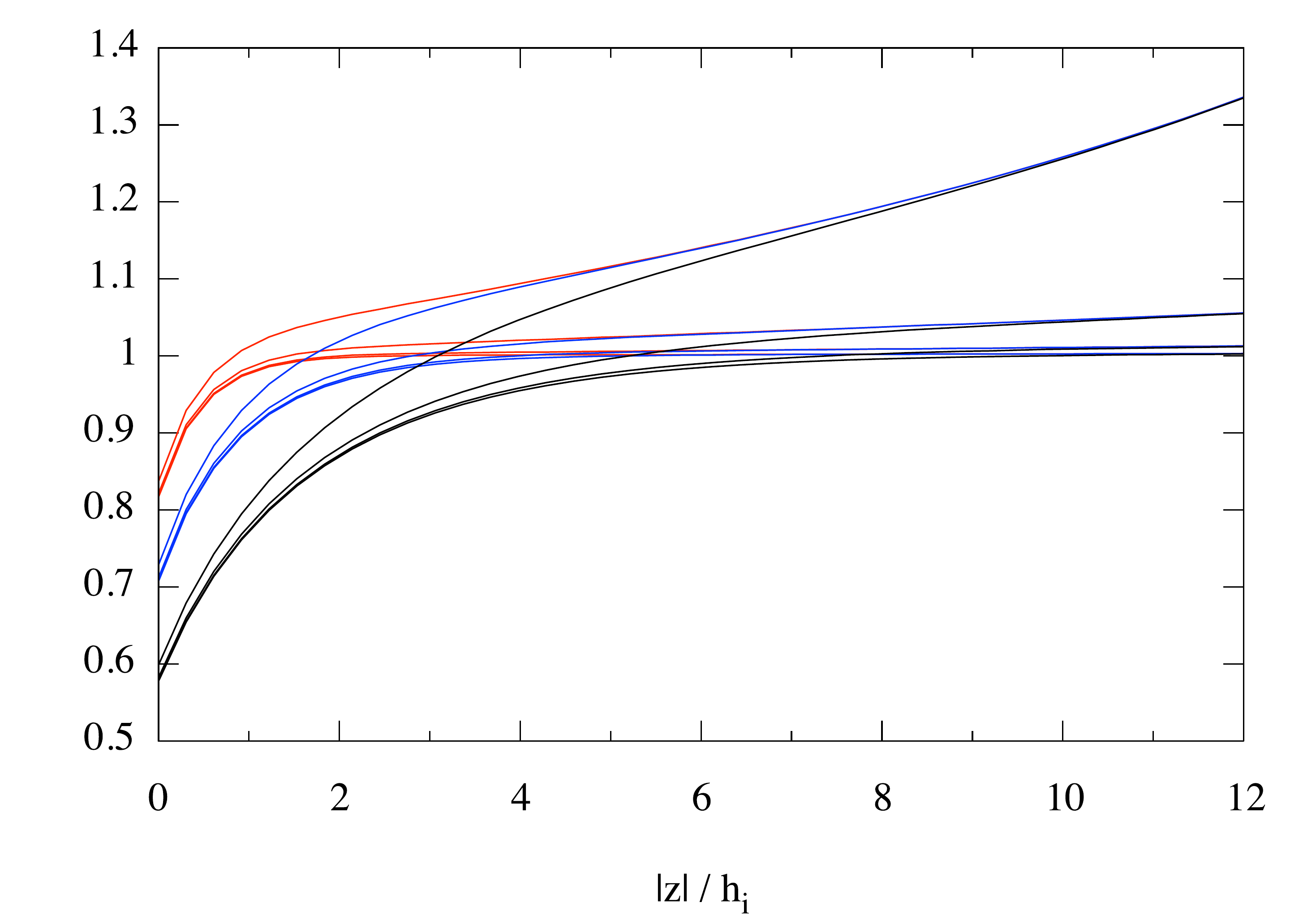}\\
\includegraphics[width=9.2cm,clip=true,trim=30   0 0 10]{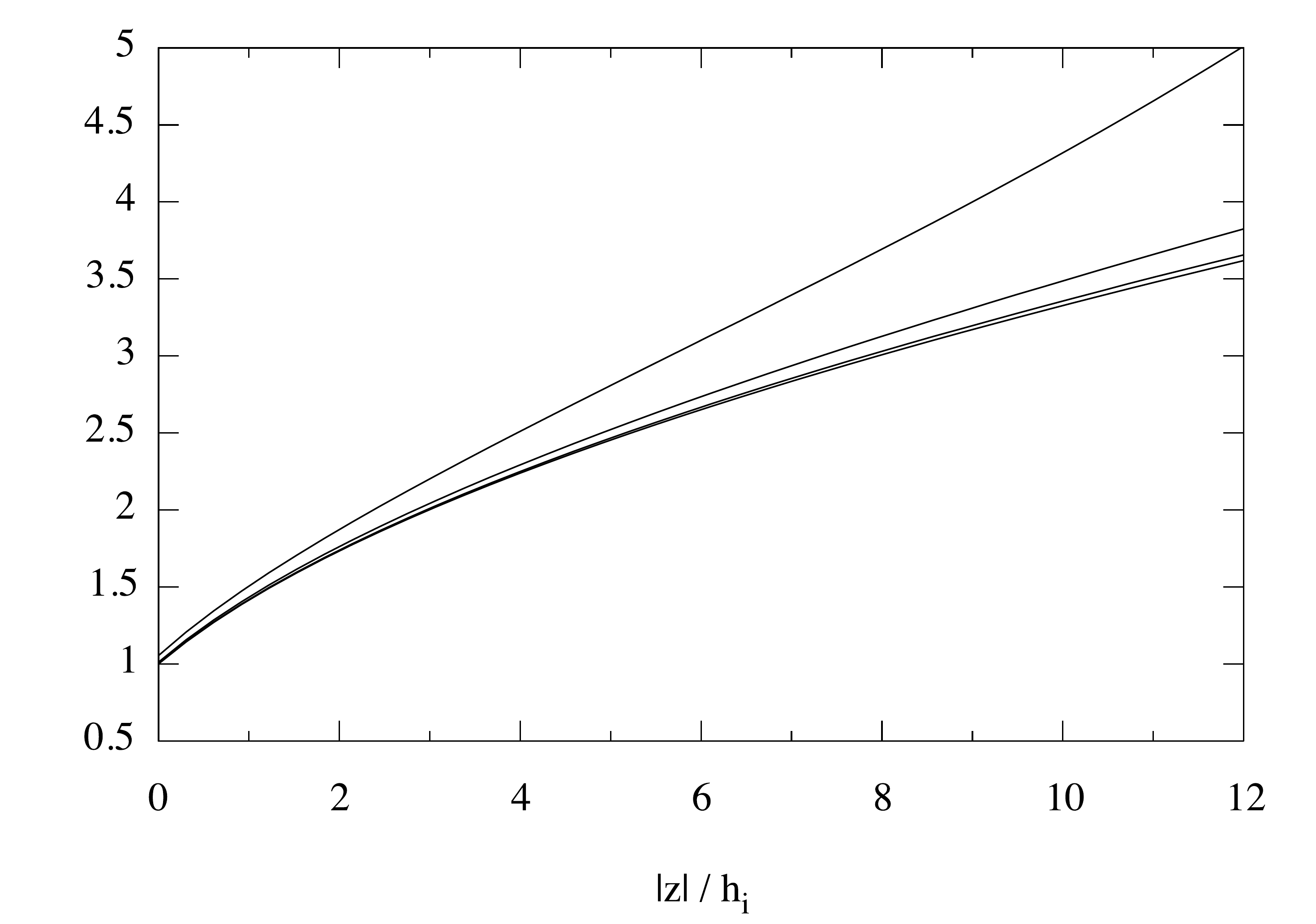}\\
\end{center}
\caption{
Top: the function $\sqrt{\Upsilon(x;h_i,\lambda_i,l_j)}$ representing the effect of a vertically exponential disc on $\sigma_{z}$ for the ratio of vertical scale-heights $l_j/h_i\!=\!0.5$ (red), 1 (blue), and 2 (black) for scaled $\sigma_{Rz}$ parameters $\lambda_i/h_i=4$ (highest), 8, 16, and 32 (lowest).
Bottom: the same for the function $\sqrt{\Psi(x;h_i,\lambda_i)}$ representing the effect of a constant density.
}
\end{figure}



\section{Thin-disc versus  thick-disc kinematics}

\label{sec:thinVSthick}

B\"udenbender et al. (\CITE{2014arXiv1407.4808B}) showed that the $\sigma_{z}(R_0,z)$ profiles of the thin- and thick-disc tracers in the SEGUE G dwarf dataset look very different  (their Fig.\,2):
the former show a large gradient of $\sim\!6\,{\KPS}{\KPC}^{-1}$ at relatively low Galactic heights ($|z| < 1.3\,\KPC$) and the latter half the gradient at much larger distances.
As emphasised by B14, the simple model for $\sigma_{z}$ used by Z13 is incapable of simultaneously explaining both the relative magnitudes and the shapes of the two curves: one expects a rapid rise in $\sigma_{z}^2$ at heights $z \! < \! h_{disc}$ to a constant value due to the disc and a subsequent linear rise due to the constant DM density (Eq.\,\ref{rhodm}).
The difference between different tracers is a simple scaling with the tracer scale-heights $h_i$ at large heights $h_i > l_j$, so the relative magnitudes are fixed by the relatively well-determined vertical density profiles; they cannot be changed at will, for example by modifying the relative amounts of baryonic and DM.
Given that the full vertical Jeans equation can now be used and that one can model the local vertical gravity more realistically, is it now possible to explain these differences, or are they simply due to tracer inhomogeneities?

At first glance, the new model would appear to be able to solve the problem entirely: different values of $\gamma_i$ (or, equivalently, of $\lambda_i$) permit either a flat or an increasing asymptotic $\sigma_z$ profile in the same disc but for different tracer populations, exactly as seen by B14.
The relative magnitudes are affected  by different values of either $\gamma_i$ or $\eta_i$, and by the difference due to relative values of the vertical scale-heights $h_i$.
In contrast to the Z13 model, a flat profile does not necessarily require unrealistic amounts of baryons, since a small exponential radial scale-length can increase the effective surface density used to calculate the gravity by large amounts.
If we find the right values for the kinematic and gravity model parameters, there would appear to be nothing in the way of determining a robust value for the DM density.

As we saw in Section\,\ref{sec:gravity}, a short radial scale-length $H \approx 2\,\KPC$ for the thick-disc means that the importance and effective scale-height of this gravity component is much larger than previously considered, placing its effects in a region where an increase in $\sigma_{z}$ was thought to be due to DM alone (Fig.\,1).
Since the gravitational effects of the thick-disc should gradually increase with height, whereas those of the thin disc eventually saturate, there should be an additional gradient in $\sigma_z$ for the thick-disc tracer data which samples the upper regions.
This is, unfortunately, exactly the opposite of what is seen in B14's two tracer samples, so the explanation is not simply a question of the gravity model.

The main kinematic parameters are $\eta_i$ and $\gamma_i$, expressible as the effective mass-density $\rho_\eta^{(i)}$ (Eq.\,\ref{rhoeta}) and the scale-length $\lambda_i$ (Eq.\,\ref{solution}). The value of
$\lambda_i^2$ is proportional to $H_i/\gamma_i$, so the smaller the value, the more significant  the effects.
While B14 did not publish $\sigma_{Rz}$ or $\sigma_{RR}$ values for their two representative thin- and thick-disc samples, one can take the kinematic data in their Table\,1 and split them into two kinematic groups, corresponding to the clearly thin discs ([$\alpha$/Fe] $< 0.2$) and clearly thick discs ([$\alpha$/Fe] $> 0.28$), i.e. the intermediate disc tracer with [Fe/H]= -0.35 and [$\alpha$/Fe]=0.28 is excluded.
This makes it possible to attempt to fit $\sigma_{Rz}$ and hence attempt to explain the differences seen by B14.

The first model (``A'') uses the tilt-angle formalism connecting $\sigma_{Rz}$ with both $\sigma_R$ and $\sigma_z$ (Eq.\,\ref{tilt-angle}).
This entails fitting the model parameters $a$ and $b$ describing $\sigma_R(\sigma_z)$ (Eq.\,\ref{sigmaR}) and then using these results to fit $\sigma_{Rz}(z,\sigma_R,\sigma_z)$ via the parameter $c$,  from which $\gamma$, $\eta$, and $\lambda$ can finally be derived
for the B14 thin, thick, and combined datasets (model ``A1'').
The fits were again made using {\tt emcee} (Foreman-Mackey et al. \CITE{2013PASP..125..306F}) following a similar fitting process to that described in Section\,2.
The value of $\chi^2_{red}$ for the thin dataset is large (3.1 vs. 1.1 for the thick dataset), but the parameter $c$ representing the tilt-angle behaviour is clearly different for the two populations ($-1.4\pm 0.2$ vs. $-2.5\pm 0.3$) unlike what one would expect for a gravitational potential representable by St\"ackel functions.
The ratio $\lambda/h$ is also different ($13\pm 3$ vs. $3.5\pm 0.5$), suggesting that the thick-disc $\sigma_z$ should have a larger gradient than the thin disc, again exactly the opposite of what is seen in B14.
Assuming that the tilt-angle behaviour is common to both tracers, one can fit a common $c$ parameter (model ``A2''); this fit shows similar values for the model parameters and the same unexpected $\lambda/h$ behaviour between the thin- and thick-disc tracers.
Since there is no reason why the $\sigma_R(\sigma_z)$-relation should be the same for all tracers, one can use an extreme model which avoids the use of $\eta$ by setting $a$ and $\eta$ equal to zero; this model (``A3'') provides adequate explanations for $\sigma_R(z,\sigma_z)$ when applied to the thin- and thick-disc data separately but results in very bad fits for the combined data ($\chi^2_\nu = 9.2$).
Despite this drastic modification of the kinematic model, the same unexpected $\lambda/h$ behaviour is seen.

The alternative phenomenological model (``B1'') is to adopt the mathematical form of $\sigma_{Rz}(z,\sigma_z),$ but to fit the two model parameters $\gamma$ and $\eta$ directly, ignoring the question of the implied effect on $\sigma_R(\sigma_z)$ within the tilt-angle assumption.
The values of $\chi^2_{red}$ for this model are lower simply because we are now fitting only the $\sigma_{Rz}$ data rather than the higher quality $\sigma_R$ data as well, but the fitted values of $\eta$ are not significant and the same $\lambda/h$ behaviour is seen.
Finally, one can fit a one-parameter model (``B2'') by assuming $\eta\equiv 0$; these fits are just as good ($\chi^2_\nu$ is 1.2 or better) and $\sigma_{Rz}$ is well fit, but one sees the same $\lambda/h$ behaviour: $\gamma$ is $1.3\pm 0.2$ and $0.8\pm 0.1$, and $\lambda/h$ is $8.9\pm 0.8$ and $3.2\pm 0.2$ for the thin- and thick-discs, respectively.

Thus, one must conclude that the differences in the kinematic behaviours of the B14 thin- and thick-disc samples are not simply due to the effects of the cross-dispersion term $\sigma_{Rz}$, at least as modelled using the tilt-angle paradigm.
The fact that a realistic gravity model  should also result in the opposite behaviour seen by B14 reinforces this conclusion.
If the effects seen in the SEGUE thin- and thick-disc data by B14 are not simply due to the inadequacies of the kinematic model, they must be due to the data.
For example, the larger gradient in the B14 $\sigma_z$ thin-disc data is mainly due to the outermost z point at $1.25\,\KPC$: if this point is lowered, then the rise at lower heights can be due to the disc scale-heights alone.
Alternatively, the effect might be due to the same behaviour seen in the MBCM data -- a gradient produced by mixing the different kinematic behaviours of different tracers.
If one compares the  B14 sampling with that of Bovy, Rix et al. (\CITE{2012ApJ...751..131B}; their Fig.\,4), one sees that the vertical scale-heights of the thin-disc mono-abundance tracers in the  B14 thin-disc samples vary from about 200 to $400\,\KPC$: using the same simple calculation made for the BT12 analysis of the MBCM data in Section\,\ref{sec:inhomogeneous}, one can produce an increase from 18 to $25\,\KPS$ over the height range of 0.40 to $1.25\,\KPC$ (Fig.\,3 in B14) with a local density ratio between the thinest and the least thin isothermal tracer component because the ratio of the relative contributions of the former over this range is roughly 10\% at the upper regions of the thin-disc data.
For a reasonable mixture of densities, one then expects an artificial gradient even in the thin-disc $\sigma_z$ data due solely to population mixture.
A similarly strong effect for the thick-disc data is not expected, since the thick-disc sample is dominated by the two most extreme thick-disc tracer populations which show roughly the same asymptotic values of $\sigma_z$, but an artificial increase in the inhomogenous vertical dispersion data must be present here as well.

If there is no problem in the vertical gradients of $\sigma_z$ measured for truly homogenous data, is there one in the relative magnitudes, as suggested by B14 (the dashed solution in their left Fig.\,2 diagram)?
The B14 ``no dark matter'' solution used a very large baryonic surface density to explain the thick-disc data, but the model overpredicted the $\sigma_z$ level for the thin disc  by about 20\%  because the asymptotic ratio $\sigma_z^2(thick)/\sigma_z^2(thin)$ is simply $h_{thick}/h_{thin}$ for the Z13 model, independent of the relative contributions of baryonic discs and the DM halo.
In the model for $\sigma_z$ presented here, the ratio also depends upon the relative values of $\lambda_i/h_i$ and $\eta_i$,
i.e. in a complicated fashion that depends upon the full details of the models and the data used.
A full discussion of these possibilities must be relegated to a subsequent paper, but it suffices to say that critical uncertainties in the vertical scale-heights and the obviously different values of $\gamma_i$ are enough to reduce this discrepancy to a level where the model is principally able to explain the observations.
Thus, the simplest explanation for the thin- and thick-disc kinematic behaviour seen by B14 is that the tracer data binned to represent only a thin and a thick disc is not homogeneous enough to permit the simple kinematic analysis used.


\section{Conclusions}

I have shown that many previous attempts at estimating the local DM density have not been as robust as traditionally perceived.
The usual baryonic corrections to the local dynamic density underestimate or ignore the uncertainties in the traditional gaseous densities and the traditional densities did not include the substantial amounts of ``dark'' gas in the form of optically thick HI and CO-dark molecular gas now known to be present in the ISM.
The non-local effects of Galactic structure increase the effective dynamic density by another 25-35\%, resulting in a total dynamic baryonic density of $120\!-160\,\mMSUNCPC$, leaving practically no room for measuring DM using mid-plane kinematics alone.
The mixing of different tracer populations with different vertical scale-heights, radial scale-lengths, and dispersion magnitudes produces gradients in $\sigma_z(z)$ that can be misinterpreted as being due to DM whenever the effects of kinematic inhomogeneity are not properly taken into account.
The gravitational effects of doubly exponential discs are quite different in form and magnitude from the infinite discs often used, particularly for radial scale-lengths $H\!<\!3\,\KPC$.
Fortunately, one can easily model this effect using additional infinite disc components with fitted surface densities and scale-heights.
For a simple model of the MW thin and thick discs, the increase in the effective total gravitating surface density is quite large,  $\sim$40\%.

Assuming only axisymmetry, stationarity, and planar symmetry, it is possible to derive a full analytic solution of the vertical Jeans equation for homogeneous tracers, a reasonably realistic local Milky Way disc plus a constant DM density that removes many of the approximations present in previous analyses.
The dispersion cross-term $\sigma_{Rz}$ can be been modelled using a phenomenologically justified expression motivated by, but not dependent upon, a tilt-angle assumption.
The additional dependence on $\sigma_z$ in the vertical Jeans equation can induce tracer-dependent and detectable additional vertical dispersion gradients that can systematically reduce the amount of {\it \emph{both}} baryonic {\it \emph{and}} dark matter needed to explain the observations.

The $\sigma_{Rz}$ model parameters were estimated using the SEQUE G dwarf data analysed by B12 and B14, either by assuming a tilt-angle model or by fitting a simple phenomenological model dependent on a linear function in $\sigma_z^2$ and a factor of $z/R$.
Despite the additional complexity of form available, it is not possible to explain the different behaviour of the ``thin'' and ``thick'' tracer data of B14 using kinematical effects alone: while the cross-dispersion effects should be large enough to be detected in the vertical profile of $\sigma_z(R_0,z)$, the changes are opposite to those needed to explain the different dispersion behaviours in the two B14 samples.
The simplest explanation is that these two inhomogeneous (non-mono-abundance) samples also show the tracer population mixture effects plaguing other analyses.

Thus, an analysis of local stellar dispersion data to extract a robust local value of $\rho_{DM}$ should in fact be possible, but it would require considerably more care and better data than has been used to date (e.g. GAIA).
In a subsequent paper, I will apply the techniques outlined herein to carefully chosen data in order to estimate the local DM density.


\begin{acknowledgements}
I would like to thank J. Bovy, A. B\"udenbender, and L. Zhang for kind access to their data.
I particularly appreciate detailed comments by C. Moni Bidin and the anonymous referee for their helpful comments that led to considerable improvements in this manuscript.
This work used the HI galaxy images available at {\tt http://www.mpia-hd.mpg.de/THINGS/Overview.html} courtesy of the THINGS collaboration.  
\end{acknowledgements}


\bibliographystyle{adsaa}
\bibliography{26022}


\end{document}